\documentclass[journal]{vgtc}                

\ifpdf
  \pdfoutput=1\relax                   
  \usepackage{graphicx}                
  \DeclareGraphicsExtensions{.pdf,.png,.jpg,.jpeg} 
\else
  \usepackage{graphicx}                
  \DeclareGraphicsExtensions{.eps}     
\fi%

\graphicspath{{figures/}{pictures/}{images/}{./}} 

\usepackage{microtype}                 
\PassOptionsToPackage{warn}{textcomp}  
\usepackage{textcomp}                  
\usepackage{mathptmx}                  
\usepackage{times}                     
\usepackage{cite}               
\usepackage{dblfloatfix}    

\usepackage{tabu}                      
\usepackage{enumitem}
\usepackage{amsmath}
\usepackage{booktabs}
\usepackage[table]{xcolor}
\usepackage{listings}
\usepackage{lstautogobble}
\definecolor{light-gray}{gray}{0.9}
\definecolor{lighter}{gray}{0.97}

\usepackage[bookmarks=false]{hyperref}

\usepackage[caption=false]{subfig}     
\usepackage{multirow}

\newcommand{\bstart}[1]{\vspace{1mm} \noindent{\textbf{#1}}}
\newcommand{\bpstart}[1]{\vspace{1mm} \noindent{\textbf{#1.}}}

\onlineid{0}

\vgtccategory{Research}
\vgtcpapertype{please specify}

\title{QualDash: Adaptable Generation of Visualisation Dashboards for Healthcare Quality Improvement}

\author{Mai Elshehaly, Rebecca Randell, Matthew Brehmer, Lynn McVey, Natasha Alvarado, Chris P. Gale, \\ and Roy A. Ruddle}
\authorfooter{
\item Mai Elshehaly, Rebecca Randell, Lynn McVey and Natasha Alvarado are with the University of Bradford and the Wolfson Centre for Applied Health Research, UK.\\ E-mail: M.Elshehaly, R.Randell, L.McVey, N.Alvarado@bradford.ac.uk.
\item Matthew Brehmer is with Tableau, Seattle, Washington, United States. E-mail: mbrehmer@tableau.com.
\item Chris P. Gale and Roy A. Ruddle are with the University of Leeds, UK. E-mail: C.P.Gale, R.A.Ruddle@leeds.ac.uk
}

\abstract{  
Adapting dashboard design to different contexts of use is an open question in visualisation research.
Dashboard designers often seek to strike a balance between dashboard adaptability and ease-of-use, and in hospitals challenges arise from the vast diversity of key metrics, data models and  users involved at different organizational levels.
In this design study, we present QualDash, a dashboard generation engine that allows for the dynamic configuration and deployment of visualisation dashboards for healthcare quality improvement (QI).
We present a rigorous task analysis based on interviews with healthcare professionals, a co-design workshop and a series of one-on-one meetings with front line analysts. 
From these activities we define a metric card metaphor as a unit of visual analysis in healthcare QI, using this concept as a building block for generating highly adaptable dashboards, and leading to the design of a Metric Specification Structure (MSS).
Each MSS is a JSON structure which enables dashboard authors to concisely configure unit-specific variants of a metric card, while offloading common patterns that are shared across cards to be preset by the engine. 
We reflect on deploying and iterating the design of QualDash in cardiology wards and pediatric intensive care units of five NHS hospitals. 
Finally, we report evaluation results that demonstrate the adaptability, ease-of-use and usefulness of QualDash in a real-world scenario. 

} 

\keywords{Information visualisation, task analysis, co-design, dashboards, design study, healthcare.}

\CCScatlist{ 
 \CCScat{K.6.1}{Management of Computing and Information Systems}%
{Project and People Management}{Life Cycle};
 \CCScat{K.7.m}{The Computing Profession}{Miscellaneous}{Ethics}
}

\teaser{
 \centering 
 \includegraphics[width=\linewidth]{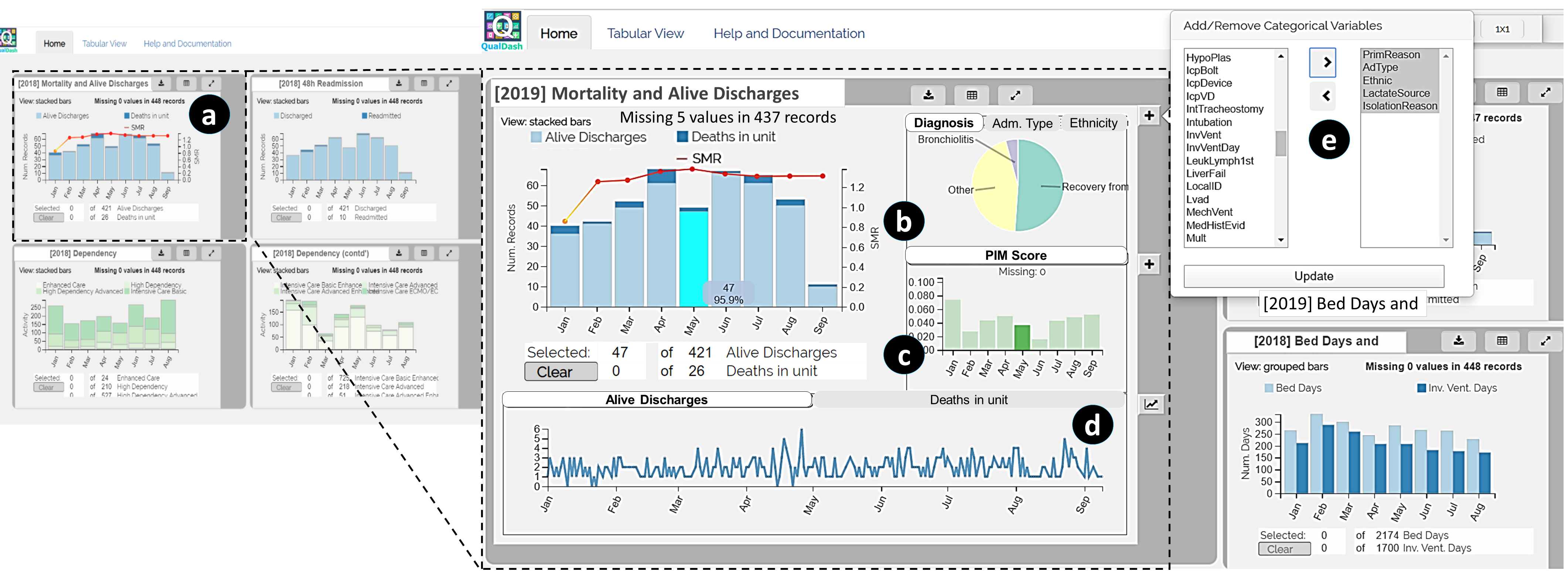}
 \vspace{-7mm}
 \caption{A dashboard with four dynamically generated QualCards (left) including one for the \textsl{Mortality} metric (a); and an expansion of the \textsl{Mortality} QualCard with categorical (b), quantitative (c) and temporal (d) subsidiary views, which are customisable via a popover (e).} 
 \label{fig:design}
}

\vgtcinsertpkg

\begin{document}


\firstsection{Introduction}
\maketitle
Visualisation dashboards are widely adopted by organizations and individuals to support data-driven situational awareness and decision making. 
Despite their ubiquity, dashboards present several challenges to visualisation design
as they aim to fulfill the data understanding needs of a diverse user population with varying degrees of visualisation literacy. 
Co-designing dashboards for quality improvement (QI) in healthcare presents an additional set of challenges due to the vast diversity of: $(a)$ performance metrics used for QI in specialised units, $(b)$ data models underlying different auditing procedures, and $(c)$ user classes involved. 
For example, while cardiologists in a large teaching hospital may monitor in-hospital delays to reperfusion treatment, some district general hospitals rarely offer this service, so they prioritise and monitor a different set of metrics. This within-specialty heterogeneity of tasks is further amplified when talking to healthcare professionals from different specialties, who use entirely different audit databases to record and monitor performance.
Consequently, the problem of designing a dashboard that can adapt to this diversity requires a high level of human involvement and  the appropriate level of automation for dashboard adaptation remains to be an open research question~\cite{sarikaya2019we}.


We present QualDash: a dashboard generation engine which aims to simplify the dashboard adaptation process through the use of customizable templates. Driven by the space of analytical tasks in healthcare QI,
we define a template in the form of a Metric Specification Structure (MSS), a JavaScript Object Notation (JSON) structure that concisely describes visualisation views catering to task sequences pertaining to individual metrics. 
The QualDash engine accepts as input an array of MSSs and generates the corresponding number of visualisation containers on the dashboard. 
We use a card metaphor to display these containers in a rearrangeable and adaptable manner, and call each such container a ``QualCard''. 
Figure~\ref{fig:design} (left) shows an example dashboard with four generated QualCards. 

\begin{figure*}
    \centering
    \includegraphics[width=.8\linewidth, height=32mm]{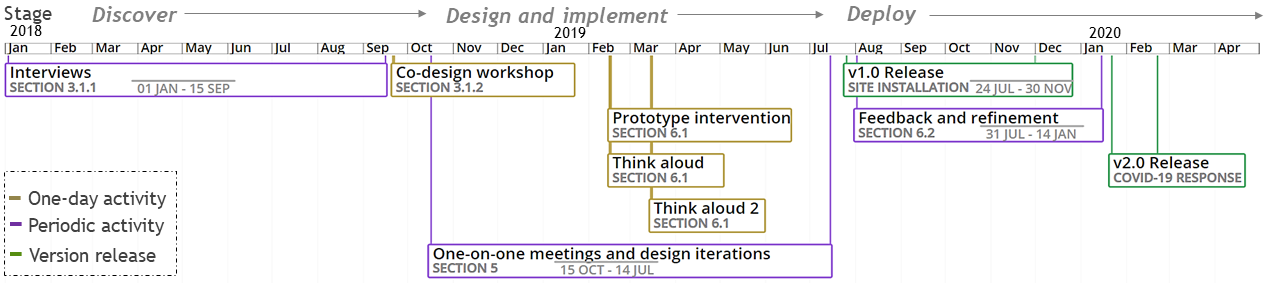}
    \caption{Timeline of the QualDash project. Design and implementation activities which spanned a period of time are given a date range and a grey bar. }
    \label{fig:outline}
    \vspace{-3mm}
\end{figure*}

We followed Sedlmair et al.'s design study methodology~\cite{sedlmair2012design} to design, implement and deploy QualDash in five NHS hospitals (Figure~\ref{fig:outline}).  
In the \textit{Discover} stage, we conducted 54 interviews with healthcare professionals and a co-design workshop, and the data we collected allowed us to characterise task sequences in healthcare QI. 
  In the \textit{Design} and \textit{Implement} stages, we conducted nine one-on-one meetings with front line analysts, as we iterated to design an initial set of key-value pairs for MSS configuration. 
Next, we conducted a second workshop, where we elicited stakeholders' intervention with a paper prototype and a high-fidelity software prototype. 
The paper prototype was designed to focus stakeholders' attention on the match between tasks and QualCards. 
The software prototype was accompanied with a think-aloud protocol to evaluate the usability of the dashboards. 
These activities resulted in an additional set of metric features, which we added in another design iteration. 
In the \textit{Deploy} stage, we deployed QualDash in five hospitals, conducted a series of ten meetings with clinical and IT staff to further adapt the MSSs to newly arising tasks, and we collected evidence of the tool's usefulness and adaptability. Suggestions for refinement were also gathered and addressed in a second version of QualDash.  

Our contributions in this paper are: 
(a) a thorough task characterisation which led to the identification of a common structure for sequences of user tasks in healthcare QI  (Section~\ref{sec:tschar});
(b) a mapping of the identified task structure to a metric card metaphor (a.k.a. the QualCard) and a Metric Specification Structure (MSS) that  allows for concise configuration of dashboards;  
(c) a dashboard generation engine that accepts an array of MSSs and generates the corresponding QualCards with GUI elements that support further customisation (Section~\ref{sec:design});  
(d) Our reflection on 62 hours of observing the deployment and adaptation of QualDash in the five NHS hospitals (Section~\ref{sec:deploy}). 




\section{Background and Related Work}
Originally derived from the concept of {balanced scorecards}~\cite{kaplan1992},  quality dashboards inherit factors that are crucial to successful adoption,
including scalability, flexibility of customisation, communication, data presentation, and a structural understanding of department-level and organization-level performance objectives and measures~\cite{assiri2006profit}. 
This paper aims to connect the dots between these factors through 
a continuous workflow that maps user \textbf{tasks} to \textbf{dashboard} specification. 
This section defines these relevant terms and outlines related work. 

\bstart{Tasks} are defined as domain- and interface-agnostic operations performed by users~\cite{Munzner2009}.
A \textit{task space} is a design space~\cite{jones2008teaching} that aims to consolidate taxonomies and typologies as a means to reason about all the possible ways  that tasks may manifest~\cite{schulz2013design}.
This concept helps visualisation researchers to \textit{``reason about similarities and differences between tasks''}~\cite{munzner2014visualization}. 
Several task taxonomies, typologies, and spaces aim to map  domain-specific tasks into a set of abstract ones that can guide visualisation design and evaluation~\cite{andrienko2006exploratory, meyer2015nested,brehmer2013multi, amar1,kerracher2017constructing, ahn2014task,schulz2013design}. 
These classifications have proven beneficial in several steps of the generative phases of visualisation design~\cite{ahn2014task, amar1,  kerracher2017constructing, sedig2013interaction, wehrend1990problem, elshehaly2018taxonomy}. 
Amar and Stasko~\cite{amar1} and Sedig and Parsons~\cite{sedig2013interaction} promoted the benefits of  typologies as a systematic basis for thinking about design.
Heer and Shneiderman~\cite{heer2012interactive} used them as constructs when considering alternative view specifications; and Kerracher and Kennedy~\cite{kerracher2017constructing} promoted their usefulness as ``checklists'' of items to consider.
By using a task space in the generative phase of design, 
 Ahn et al.~\cite{ahn2014task} identified previously unconsidered tasks in network evolution analysis.
We leverage the opportunities that task classification offers to support our understanding of common structures for task sequences in the context of dashboard design in healthcare QI.

\bstart{Dashboards} are broadly defined as \textit{``a visual display of data used to monitor conditions and/or
facilitate understanding''}~\cite{wexler2017big}.
Sarikaya et al. highlighted the variety of definitions that exist for dashboards
~\cite{sarikaya2019we}, while Few highlighted the disparity of information that dashboard users typically need to monitor~\cite{few2006common}.
We focus our attention on the definition and use of visualisation dashboards in a healthcare context, where Dowding et al. 
 distinguished two main categories of dashboards that inform performance~\cite{dowding2015dashboards}:
 \textbf{clinical dashboards}  provide clinicians with timely and relevant feedback on their patients' outcomes,  
 while 
\textbf{quality dashboards} are meant to inform \textit{``on standardized performance metrics at a
unit or organizational level''}~\cite{dowding2015dashboards}. 
Unlike clinical dashboards which cater to a specialized user group within a specific clinical context (e.g.~\cite{idmvis, composer, loorak2015timespan}), we further set the focus on quality dashboards, which exhibit a wider variety of users,  contexts, and tasks.

\bstart{The healthcare visualisation literature} presents tools that are broadly classified~\cite{rind2013interactive} into ones that focus on individual patient records (e.g., LifeLines~\cite{milash1996lifelines}), and ones which display aggregations (e.g. Lifelines2~\cite{wang2010interactive}, LifeFlow~\cite{wongsuphasawat2011lifeflow}, EventFlow~\cite{6634100}, DecisionFlow~\cite{6875996} and TimeSpan~\cite{loorak2015timespan}). 
A common theme is that these tools dedicate fixed screen real-estate to facets of data.
Consequently, they support specialised tasks that focus on specific types of events (e.g. ICU admissions~\cite{wang2010interactive}), or specific patient cohorts (e.g. stroke patients~\cite{loorak2015timespan}).  

\bstart{Commercial software} such as Tableau offer a wealth of expressivity  for dashboard generation, allowing interactive dashboards to be deployed to members of an organisation \textit{``without limiting them to pre-defined questions''}~\cite{rueter2012tableau}.
Similar levels of expressivity are also attainable with grammars such as Vega~\cite{satyanarayan2015reactive} and Vega-lite~\cite{satyanarayan2017vega}.
These grammars empower users' to explore many visualisation alternatives, particularly when encapsulated in interactive tools~\cite{wongsuphasawat2015voyager,wongsuphasawat2017voyager}. 
More recently, Draco~\cite{8440847} subsets Vega-lite's design space to achieve an even more concise specification. 
However, this specification and its precedents do not offer a mechanism to encode users' \textit{task sequences}. 
In contrast, we contribute an engine that leverages taxonomic similarities of identified tasks to  present ``templates'' for dashboard view specifications. 
While our MSS spans a subset of the design space of the aforementioned tools, we show that our concise templates enable the co-design and deployment of dashboards in healthcare QI. 
To our knowledge, this paper presents the first attempt to capture task sequences for audiences in healthcare QI and to match this characterisation to dynamic dashboard generation.

\section{Task analysis}
\label{sec:tschar}
\begin{figure*}[h]
 \centering 
     \subfloat[\label{left}]{
        \includegraphics[width=.4\linewidth, height=1.6in]{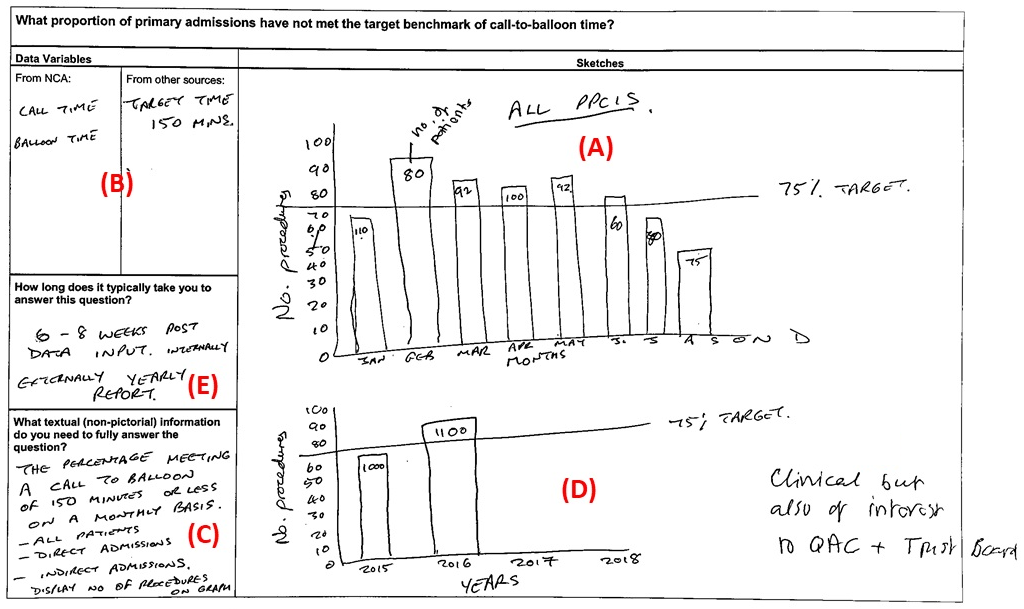}
    }
    \hfill
    \subfloat[\label{left}]{
        \includegraphics[width=.28\linewidth, height=1.55in]{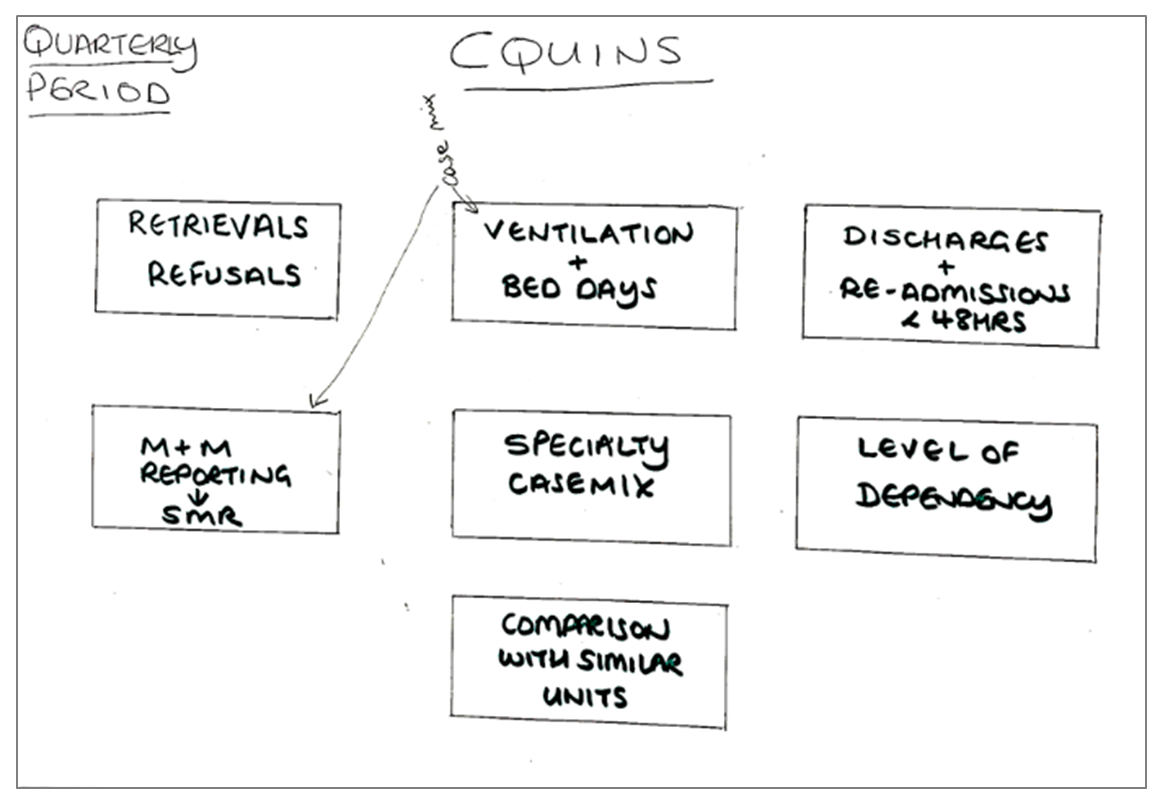}
    }
    \hfill
    \subfloat[\label{right}]{
        \includegraphics[width=.28\linewidth, height=1.7in]{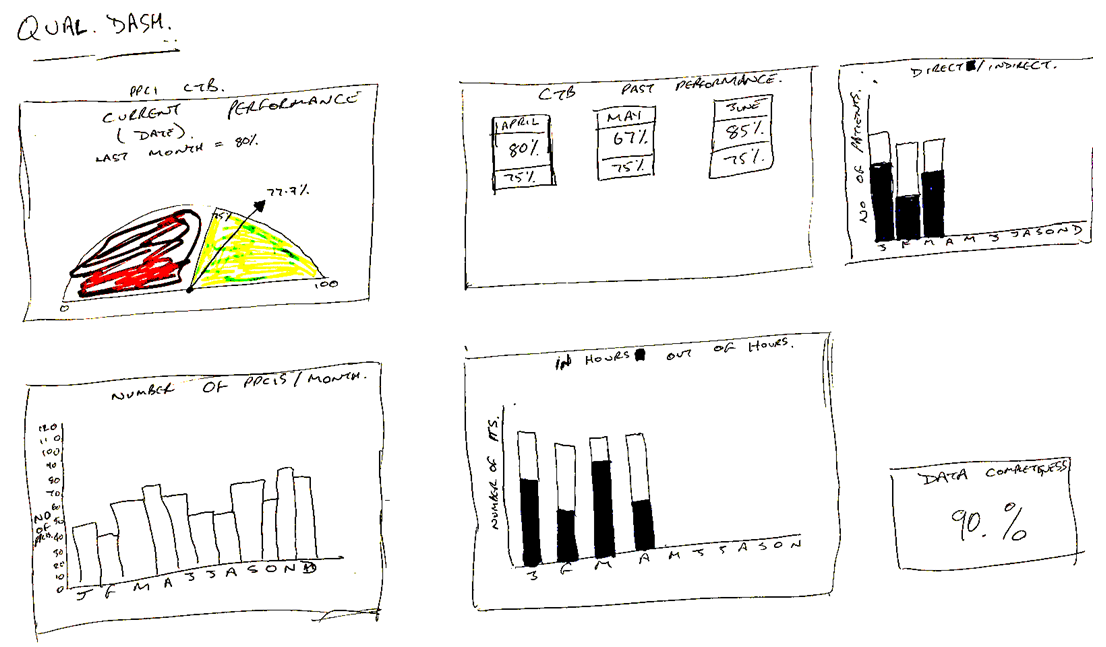}
    }
 \vspace{-2mm} 
 \caption{Examples of participants' responses to Co-design Activity 1 (Story Generation) (a), and Co-design Activity 2 (Task Sequencing) (b and c). } 
 
 \label{fig:activ1}
\end{figure*}

Our analysis is guided by the dimensions of the dashboard design space identified by 
Sarikaya et al.~\cite{sarikaya2019we} and the challenges specific to healthcare QI~\cite{Randelle033208}. 
Namely, we sought to answer the questions: What user task sequences  exist within and across audiences of healthcare QI?
How are metrics and benchmarks defined? 
    What visual features strike a balance between ease-of-use and adaptation?
    And how updatable should the dashboards be? 


\subsection{Data Collection}
Our data collection started with an investigation of challenges and opportunities inherent in the use of National Clinical Audit (NCA) data for healthcare QI. 
NCAs provide national records of data about patient treatments and outcomes in different clinical areas.  
Participating hospitals regularly contribute to updates of over 50 audits, to systematically record and measure the quality delivered by clinical teams and healthcare organizations. 
We focus on two audits: the Paediatric Intensive Care Audit Network (PICANet) audit~\cite{picanet} and the Myocardial Ischaemia National Audit Project (MINAP) audit~\cite{wilkinson2020myocardial}.
Both audits provided ample data to work with as well as access to a diversity of providers and stakeholders.
We characterised the space of QI task sequences through interviews and a co-design workshop.

\subsubsection{Interviews with Stakeholders}
\label{sub:interview}
We interviewed 54 healthcare professionals of various backgrounds, including 
physicians, 
nurses,  
support staff,  
board members, 
quality \& safety staff, 
and information staff. 
The interview questions elicited the participants’ own experiences with NCA data, with an emphasis on the role of the
data in informing quality improvement. Interviewees gave examples of when audit data were of particular
use to them, where limitations and constraints arose, and what
they aimed to learn from the data in the future. 
All interviews were audio-recorded and transcribed verbatim. 
We used these qualitative data to build a thematic understanding of audit-specific quality metrics. 
For each metric, we collated queries and quotes from the interview transcripts that led to the generation of an initial pool of 124 user tasks (Appendix A). 

The next step was to establish context around these tasks, and to identify connecting sequences among them.  
Since including all 124 tasks for this contextualisation was not feasible, we selected a representative subset of tasks by considering structural characteristics, based on the three taxonomic dimensions highlighted for healthcare QI: granularity, type cardinality and target~\cite{elshehaly2018taxonomy}.
When selecting a reduced task subset, we
included tasks that covered different granularity (e.g., unit or national level of detail)  and type-cardinality levels (i.e. number of quantitative, temporal and  categorical variables) from the task pool. 

\subsubsection{Co-design Workshop}

We organised a half-day co-design workshop to build task sequences and contexts. 
Workshop participants were seven staff members from one hospital having clinical and non-clinical backgrounds (2 consultants, 2 nurses and 3 hospital information managers). 
We divided participants into two groups, depending on which audit they were most familiar with and assigned each group a task set that corresponds to one of the audits.
Two project team members facilitated the discussion with each group over the course of two co-design activities. 

\bpstart{Co-design Activity 1: Story Generation}
Inspired by Brehmer and Munzner's approach to task analysis~\cite{brehmer2013multi}, we asked participants to answer questions about:
\textit{why} tasks are important, 
\textit{how} tasks are currently performed, and
\textit{what} information needs to be available to perform a task (\textit{input}) as well as \textit{what} information might arise after completing the task (\textit{output)}.
We added a fourth question: \textit{who} can benefit most from a task, which is inspired by agile user story-generation~\cite{cohn2004user}.

To facilitate the discussion around these four questions, co-design participants were presented with a set of ``task cards'' (Figure~\ref{fig:activ1}a). 
Each card focused on a single task and was subdivided into four main sections used to collect contextual information about it.
The header of a card contained the task's body and an empty box for participants to assign a relevance score. Three parts of the card were dedicated to elicit information about how this task is performed in current practice: the data elements used, the time it takes, and any textual information that might be involved (sketch areas (B), (E) and (C), respectively, in Figure~\ref{fig:activ1}$a$).
The majority of space on the card was dedicated to a large ``sketches'' section. This section provided a space for participants to sketch out the processes involved in performing the task, as well as any visualisations used in the same context. 

Participants were presented with a set of task cards corresponding to the reduced task space described in Section~\ref{sub:interview}. 
We also gave each participant a set of blank cards, containing no task in the task body section.
Participants were given the freedom to select task cards that they deemed relevant to their current practice, or to write their own. 
For each task, participants were asked to solve the \textit{what} and \textit{how} questions individually, while the \textit{why} and \textit{who} questions were reserved for a later group discussion. 
During the discussion, we asked participants to justify the relevance scores they assigned to each task (\textit{why}), elaborate on their answers, and then sort the cards depending on \textit{who} they believed this task was most relevant to. 

\bpstart{Co-design Activity 2: Task Sequencing}
From the tasks that were prioritised in Activity 1, we asked participants to select \textit{entry-point} tasks. These are questions that need to be answered at a glance without interacting with the dashboard interface. Once completed, we explained, these tasks may lead to a sequence of follow-up tasks. 
To identify these sequences, we returned the prioritised task cards from Activity 1 to participants. 
We asked participants to: 
    \textit{(i)} Select the most pressing questions to be answered at a glance in a dashboard;
    \textit{(ii)} Sketch the layout of a static dashboard that could provide the minimally sufficient information for answering these tasks; and
    \textit{(iii)} Select or add follow-up tasks  that arise from these entry-point tasks.
\subsubsection{Structure of Task Sequences}
Our activities revealed that the use of NCA data is largely at
the clinical team level, with more limited use at divisional and corporate levels. 
We identified entry-point tasks that required monitoring five to six key metrics for each audit (see Figure~\ref{fig:activ1}b). 
We have included a glossary of terms in supplementary material to explain each of these metrics.

\begin{table*}[h]
	\caption{Task sequence pertaining to the \textsl{call-to-Balloon} metric. Code prefixes $MEP$ and \texttt{MSUB} indicate whether a task is an entry-point or subsidiary. }
	\label{tab:mort}
	\centering
	\scriptsize
	\resizebox{0.95\textwidth}{!}{%
	\begin{tabular}{|c|l|c|c|c|c|}
		\hline
		 \textbf{Code} & \textbf{Task} & Quan. & Nom. & Ord. & Temp. \\
		\hline
		\hline
		$MEP1-1$ &  \textbf{What proportion of primary admissions have / have not met the target benchmark of call-to-balloon time? } & 2 &  & & 1 \\
		$MEP1-2$&	\multicolumn{1}{|l|}{\textbullet~ On a given month, what was the total number of PCI patients? } &  1 &  & & 1   \\
	 	$MEP1-3$&	\multicolumn{1}{|l|}{\textbullet~ On a given month, 
did the percentage of primary admissions that met the target of 150 minutes call-to-Balloon time exceed 70\%? } &  1 &  & & 1   \\
		
		\hline    
	    \texttt{MSUB1-1} & \multicolumn{1}{|l|}{\textbullet~ Of STEMI patients that did not meet the call-to-Balloon target, which ones were direct/indirect admissions?  } &  & + 1 & &  \\
		\texttt{MSUB1-2} & \multicolumn{1}{|l|}{\textbullet~ Where did the patients that did not meet the target come from? } &  & + 1  & &   \\
		\texttt{MSUB1-3} & \multicolumn{1}{|l|}{\textbullet~ Are delays justified? } &  & + 1  & &   \\
	 \hline
		\texttt{MSUB1-4} & \multicolumn{1}{|l|}{\textbullet~ How does a month with high number of PCI patients compare to the same month last year? } &  &  & & + 1  \\
		 \hline
	    \texttt{MSUB1-4} & \multicolumn{1}{|l|}{\textbullet~ Did the patients who did not meet the target commute from a far away district?  } &  &  & + 1 &  \\
	     \hline
	    \texttt{MSUB1-5} & \multicolumn{1}{|l|}{\textbullet~  For a month with a high number of delays, what was the average door-to-balloon time?  } & + 1  & & &  \\
		\hline
			\texttt{MSUB1-6} & \multicolumn{1}{|l|}{\textbullet~ What is the life and death status of delayed patients 30 days after leaving the hospital? } &  \multicolumn{4}{c|}{Excluded}    \\
		 \texttt{MSUB1-7} & \multicolumn{1}{|l|}{\textbullet~ Compare the average of cases meeting the call-to-balloon target for own site versus district  } &  \multicolumn{4}{c|}{Excluded} \\ 
	    \hline
	\end{tabular}
	}
\end{table*}

Our analysis led to three key findings:  
\textbf{[F1]} individual metrics have independent task sequences;
\textbf{[F2]} each metric has entry-point tasks that involve monitoring a small number of measures over time; and 
\textbf{[F3]} investigation of further detail involves one or more of three abstract {\it subsidiary tasks}: 
\begin{itemize}[noitemsep,nolistsep,leftmargin=*]
\item \texttt{ST1}: Break down the main measure(s) for patient \texttt{\textbf{sub-categories} }
    \item \texttt{ST2}: Link with other metric-related \texttt{\textbf{measures} }
    \item \texttt{ST3}: Expand in \texttt{\textbf{time}} to include different temporal granularities
\end{itemize}

\textbf{[F1]} was noted by participants during Activity 1 and maintained through Activity 2. 
Figure~\ref{fig:activ1} shows example responses for different audits.
In Figure~\ref{fig:activ1}$b$, a participant explained that a dashboard should provide a minimalist entry point into the metrics of interest to their Pediatric Intensive Care Unit (PICU). Another participant advocated this design by saying: \textit{“I want something simple that tells me where
something is worsening in a metric, then I can click and find out more”}.

In Figures~\ref{fig:activ1}$a$ and $c$, participants faceted views for sequences pertaining to the call-to-balloon metric, for example. 
They explained that for this metric, patients diagnosed with ST Elevation Myocardial Infarction (STEMI) - a serious type of heart attack - must have a PPCI (i.e. a balloon stent) within the national target time of 150 minutes from the time of calling for help. 
An entry-point task  for this metric regards monthly aggregates of admitted STEMI patients, and the ratio who met this target (Figure~\ref{fig:activ1}$a$ sketch area (A), and Figure~\ref{fig:activ1}$c$ bottom left).
Participants then linked this to a breakdown of known causes of delay to decide whether they were justified (\texttt{ST1}). 
One source of delay, for instance, may be if the patient was self-admitted. 
This information was added by a participant as textual information in Figure~\ref{fig:activ1}$a$ (sketch area (C)) and by another participant as a bar chart (top right corner of Figure~\ref{fig:activ1}$c$).
Participants also noted that it is important to investigate the measures in a historic context (\texttt{ST3}) by including previous months (Figure~\ref{fig:activ1}$c$ top middle) and years (Figure~\ref{fig:activ1}$a$ sketch area (D)).

Table~\ref{tab:mort} lists the tasks of the call-to-balloon metric along with counts of different types of data in each task, as defined in~\cite{satyanarayan2017vega}. $Quan$tiative, $Nom$inal, $Ord$inal and $Temp$oral measures required for the entry-point tasks are listed and additional measures  considered for subsidiary tasks are marked with a $+$ sign.
Despite the variability of metrics across audits, the structure of entry point and subsidiary tasks remains the same. 
Appendix B lists the task sequences we identified for all metrics.  

\section{Design Requirements for QualDash}
\label{sec:req}
Equipped with a well-defined structure of task sequences, we looked into the use of visualisation grammars like Vega-lite~\cite{satyanarayan2017vega} to generate views on a dashboard arranged to serve the identified tasks. 
Findings from the interviews and co-design workshop were further discussed in a sequence of nine one-on-one meetings with front-line analysts over the course of nine months. Front line analysts are audit coordinators and clinicians who are well-acquainted with audit data as they use it for reporting, presentation and clinical governance. 
Our meetings involved two consultant cardiologists, 
a consultant pediatrician, and two audit coordinators. Two of the consultants also held the title ``audit lead''. 

During these meetings, we presented the concept of a QualCard as a self-enclosed area of the dashboard screen that captures information pertaining to a specific metric. 
We demonstrated design iterations of QualCard prototypes and discussed key properties that would be minimally sufficient to configure them. 
We leveraged the analysts' familiarity with the data by discussing queries as we sought to specify the \texttt{field}, \texttt{type} and \texttt{aggregate} properties of a Vega-lite data axis.
This exercise helped us exclude tasks that required data that the audit did not provide (e.g., Task \texttt{MSUB1-6} in Table~\ref{tab:mort}).
We also excluded tasks that required data not accessible within individual sites (e.g., Task \texttt{MSUB1-7} in Table~\ref{tab:mort}, because comparing against other sites was deemed infeasible as it required cross-site data sharing agreements).

Next, we explored different ways to compose layered and multi-view plots within each QualCard to address the identified task sequences. 
A number of design requirements emerged from these meetings: 

\begin{itemize}[noitemsep,nolistsep]
   \item [R1] \textbf{Support pre-configured reusable queries for dynamic QualCard generation.}
    Given the variability of metrics across sites and specialties, each unit requires a dynamically-generated dashboard that captures unit-specific metrics. Pre-configuration is necessary at this level of dashboard authoring, to define care pathways that lead a patient's record to be included in a metric. 
    
    \item [R2] \textbf{Each QualCard must have two states:} 
    \begin{itemize}[noitemsep,nolistsep]
    \item [R2.1] \textbf{An entry-point state} in which a QualCard only displays the metric's main measures aggregated over time.
    \item [R2.2] \textbf{An expanded state} in which a QualCard reveals additional views, catering to subsidiary tasks \texttt{ST1}, \texttt{ST2} and \texttt{ST3}. 
   \end{itemize}
        
    \item [R3] \textbf{Support GUI-based adaptability of subsidiary view measures} to cater to different lines of inquiry.
    \item [R4] \textbf{Data timeliness}: support varying workflows and frequencies in data updates.
    \item [R5] \textbf{Data quality}: present information on missing and invalid data pertaining to a metric's specific measures. 
    \item [R6] \textbf{Support exports} of visualisations and individual data records to be used in clinical governance meetings.
    \item [R7]\textbf{ Data privacy:} data cannot leave the hospital site. 
\end{itemize}

\begin{figure}[h]
 \centering 
 \includegraphics[width=\linewidth]{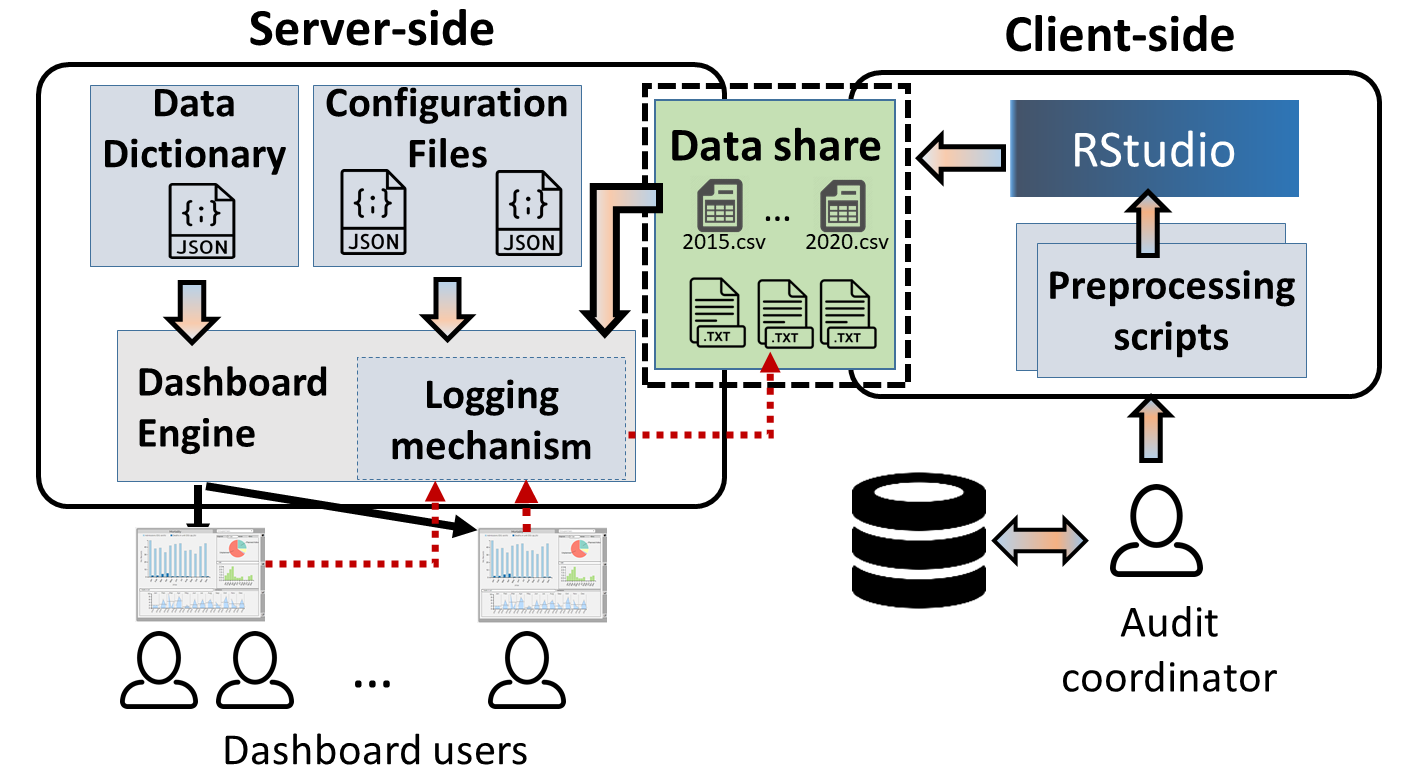}
 \caption{The QualDash client-server architecture: the dashboard engine reads an array of MSSs from configuration files and fetches audit data supplied by an audit coordinator, to generate the QualCards. } 
 \label{fig:schematic}
 \vspace{-2mm}
\end{figure}

\section{QualDash Design}
\label{sec:design}


QualDash is a web-enabled dashboard generation engine that we designed, implemented and deployed to meet the above requirements.
The QualDash architecture (Figure~\ref{fig:schematic}) consists of client-side and server-side components.
Both the client and the server are setup locally within each hospital site so that data never leaves the site (R7). 
Audit data is supplied by an audit coordinator using a client machine and kept at an on-site data share that is accessible from the server. 

To support timeliness (R4), QualDash includes an R script that performs  pre-processing steps and uploads data to the shared location.
Data pre-processing includes calculations to: (i) convert all date formats to the Portable Operating System Interface (POSIX) standard format~\cite{atlidakis2016posix} to support an operating-system-agnostic data model; 
(ii) calculate derived fields which are not readily available in audit data (e.g., we use the EMS R package~\cite{ems} to pre-calculate the risk-adjusted Standardised Mortality Ratio (SMR) measure from PICANet data); and (iii) organise audit data into separate annual files for efficient loading in a web browser.

\begin{figure*}[h]
\subfloat[\label{left}]{ \fbox{\includegraphics[width=0.4\linewidth, height=49mm]{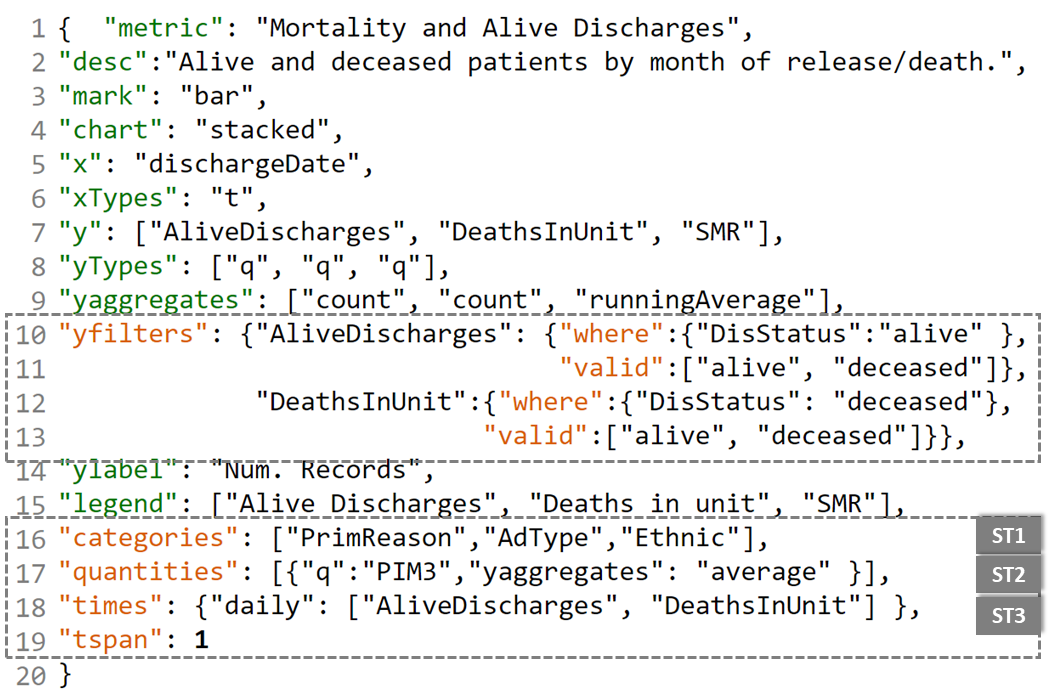}}}
  \subfloat[\label{right}]{  \fbox{\includegraphics[width=0.565\linewidth, height=49mm]{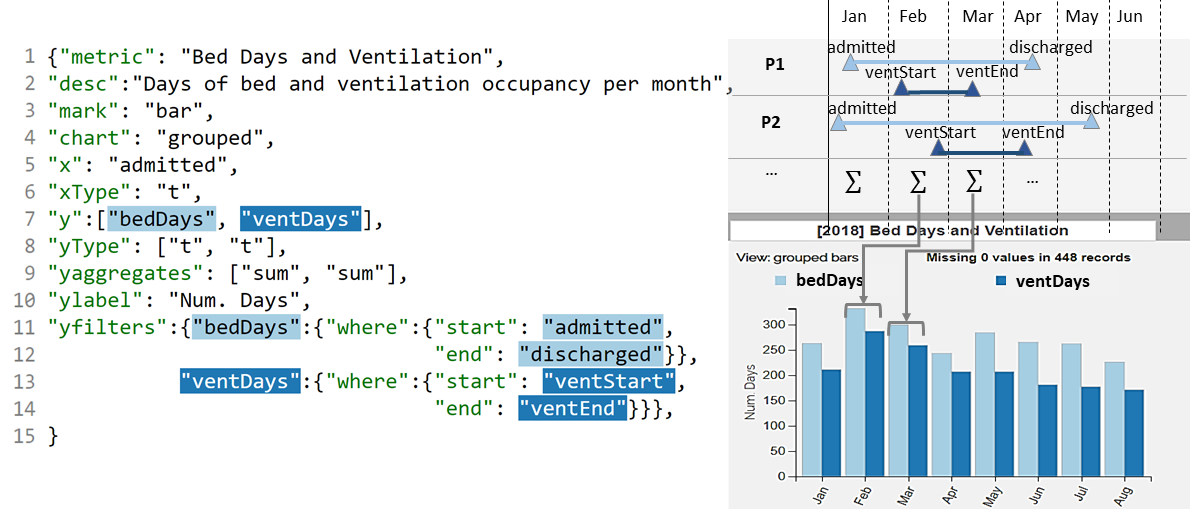}}}
    \vspace{-3mm}
    \caption{Examples of the MSS for: (a) The ``Mortality and Alive Discharges'' QualCard that is shown in Figure~\ref{fig:design}. The outlined parts manage inclusion criteria (Lines $10-13$) and subsidiary tasks (Lines $16-19$). (b) The ``Bed days and Ventilation'' QualCard includes temporal measures, derived from the \texttt{start} and \texttt{end} variables specified in \texttt{yfilters}. The QualDash engine breaks down intervals between \texttt{start} and \texttt{end} to calculate aggregates. }
    \vspace{-3mm}
    \label{fig:mss}
\end{figure*}

On the server-side, we developed a web tool that allows users to specify an audit and a time frame, and renders the corresponding dashboard in a web browser running a backend dashboard generation engine. The dashboard is dynamically generated and rendered from a JSON configuration file which resides on the server (R1). 
Configuration files supply an array of Metric Specification Structures (MSSs) to the QualDash engine, which in turn generates the corresponding QualCards.
 Each QualCard is a self-contained building block for dashboards that encompasses all entry-point and subsidiary information relating to a single metric (R2). 
A data dictionary is supplied by the audit supplier and contains natural language text descriptions of individual fields. Finally, a logging mechanism  records users' interactions. 
Anonymous usage logs (R7) are fed back to the data share so that they can be accessed by the audit coordinator who then sends us the logs.  The remainder of this section details the MSS design and describes how the dashboard engine interprets its elements to generate the QualCard interface.  


\subsection{The Metric Specification Structure (MSS)}
\label{sub:mss}

The MSS is a self-contained JSON structure that includes information to dynamically generate concatenated views and support task sequences for an individual QI metric, including main measures and subsidiary components (R1- R2). 
This only requires a subset of the expressivity of general-purpose visualisation grammars like Vega-Lite~\cite{satyanarayan2017vega}. 
Unlike Vega-Lite, which provides a large variety of possibilities for view composition via faceting, concatenation, layering and repeating views, the MSS provides a constrained design space that sets specific relationships across concatenated views, and allows us to concisely configure a QualCard; while leaving it up to the QualDash engine to set defaults that are common across metrics. 

 

Figure~\ref{fig:mss}$a$ shows an example MSS which configures the ``Mortality and Alive Discharge'' QualCard. 
The MSS defines: \textit{(i)} a metric name and a description of its main measures (Lines $1-2$); \textit{(ii)} a compact version of Vega-lite's \texttt{data}, \texttt{mark} and \texttt{encoding} fields (Lines $3-9$); \textit{(iii)} inclusion filters (Lines $10-13$); \textit{(iv)} an axis label and   view legend (Lines $14-15$); and 
\textit{(v)} information for the subsidiary views (Lines $16-19$).
Appendix D provides the specifics of each of these MSS keys and maps them to the corresponding predicates in Vega-Lite~\cite{satyanarayan2017vega}. We focus our discussion here on keys that capture the most relevant functionality for QualDash (outlined in Figure~\ref{fig:mss}$a$).

The \texttt{yfilters} key allows the specification of inclusion criteria for patients considered in each measure. In the mortality metric example shown in Figure~\ref{fig:mss}$a$, both the \texttt{AliveDischarges} and \texttt{DeathsInUnit} measures
specify filtering criteria based on the discharge status of a patient as \texttt{"alive"} and \texttt{"deceased"}, respectively.  
In cases where multiple key-value pairs are used to filter the data, we define an \texttt{operator} that combines multiple criteria (not shown in Figure~\ref{fig:mss}) using either a logical AND or OR operator. 
At the moment, the QualDash engine does not support composite criteria as this was deemed unnecessary for the majority of the healthcare QI tasks collected in our analysis. 
In the rare cases where composite criteria are required, we offload part of the composition to our R pre-processing scripts. Section~\ref{sub:gold} describes a use case that exemplifies this scenario. 

The \texttt{yfilters} key extends Vega-Lite's \texttt{Filter} transform~\cite{satyanarayan2017vega} in two ways: 
(1) To capture information on data quality (R5), we include a \texttt{valid} field that, rather than checking for 
missing data only, accepts a list of valid values for each measure. This is to accommodate audits where invalid entries are coded with special values in the data. 
(2) To support temporal aggregation, we define two special keys within the \texttt{where} clause of \texttt{yfilters}. The \texttt{start} and \texttt{end} keys specify boundary events in cases where a measure spans a period of time. For example, bed and ventilation days per month are the two main measures shown in Figure~\ref{fig:mss}$b$. 
Since each patient can occupy a bed or a ventilator for an elongated period of time, we specify two derived fields to define these measures: \texttt{bedDays} and \texttt{ventDays} (Line $7$ in Figure~\ref{fig:mss}$b$). 
The QualDash engine looks for hints in the \texttt{where} clause to derive these measures.
The dates on which a patient was \texttt{admitted} to and \texttt{discharged} from a hospital specify the \texttt{start} and \texttt{end} events of the \texttt{bedDays} measure, respectively.
The QualDash engine calculates the days elapsed  between these two events, and aggregates the corresponding monthly bins.  
Similarly, the  \texttt{ventDays} measure is calculated using the time elapsed between the \texttt{ventStart} and \texttt{ventEnd} events.

For a QualCard's subsidiary views, a \texttt{categories} field defines what categorical variables are used to break down the main measures of the metric (\texttt{ST1}). For example, when clinicians regard patient mortality in a Pediatric Intensive Care Unit (see Line $16$ in Figure~\ref{fig:mss}$a$), they consider a breakdown of the primary reason for admission (\texttt{PrimReason}). They also check the admission type  (\texttt{AdType}); and investigate whether a specific ethnicity had a higher mortality rate (\texttt{Ethnic}). 
A \texttt{quantities} field (Line $17$ in Figure~\ref{fig:mss}$a$) captures additional measures that users link to the main measures and would want to understand within the same context (\texttt{ST2}). 
The \texttt{quantities} are defined using the variable name(s) and an aggregation rule that is to be applied. 
For both the main view and the \texttt{quantities} view, the \texttt{yaggregates} key supports \texttt{count}, \texttt{sum},  \texttt{runningSum}, \texttt{average} and \texttt{runningAverage} aggregation rules. 
Finally, the \texttt{times} field (line $18$ in Figure~\ref{fig:mss}$a$) specifies a default temporal granularity for additional historical context (\texttt{ST3}). This field accepts a key-value pair where the key describes the time unit to be displayed by default, and the value lists measures that need to be displayed at this granularity. 
The number of years included in this temporal context is defined by the \texttt{tspan} field (line $19$ in Figure~\ref{fig:mss}$a$).

The MSS keys described in this section allow dashboard authors to ensure its safe use and interpretation (see Section~\ref{sub:gold}) by capturing definitions that lead to a patient's inclusion in a measure. They also capture known task sequences that were identified in the analysis phase. 



\subsection{The QualCard Interface}
\label{sub:card}


\begin{figure*}[h]
    \centering
    \includegraphics[width=.94\linewidth]{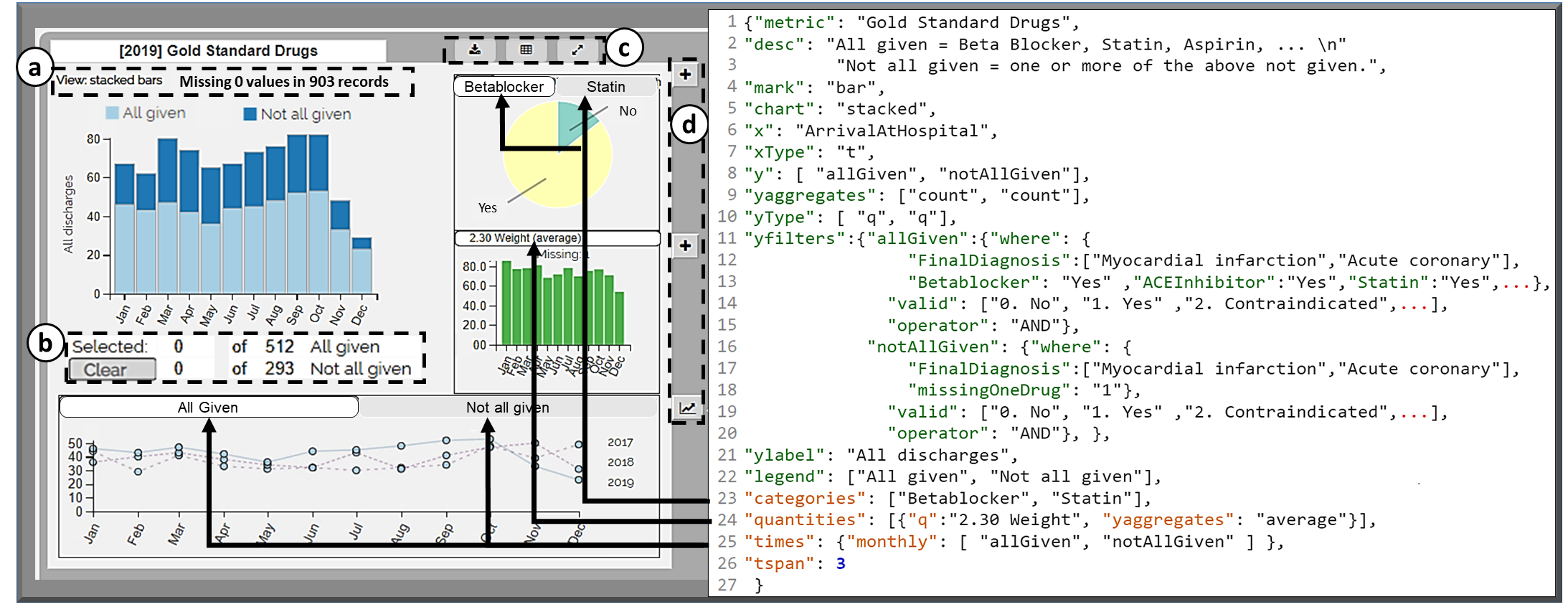}
    
    \caption{QualCard for the Gold Standard Drugs metric (left) and the corresponding MSS (right). (a) information on data missingness (derived from lines $14$ and $19$ in the MSS); (b) selection information for each measure; (c) control panel; and (d) sub-view customisation buttons. }
    \vspace{-2mm}
    \label{fig:gold}
\end{figure*}

The purpose of a QualCard is to act as a self-enclosed reusable and dynamically-generated visualisation container which captures task sequences pertaining to a single QI metric. 
To design this card metaphor, we first introduced our front-line analysts to the idea of a \textit{Metric Bubble}, inspired by Code Bubbles~\cite{bragdon2010code} and VisBubbles~\cite{li2011visbubbles}, as enclosed views that allow flexible dragging and resizing.
The idea was well-received. However, the free dragging behaviour was perceived as overwhelming. 
One clinician argued for a more restricted mechanism. 
In response, we considered two layout alternatives: a tabbed view and a grid view~\cite{10.1145/2213836.2213931}.
Empirical evidence shows that juxtaposed
patches on a grid layout significantly reduce navigation mistakes~\cite{henley2017toward}.
Additionally, input from co-design showed an expectation by stakeholders to be able to obtain an overview of all metrics on the same screen at the entry point of their analysis.
We therefore opted to display QualCards on a grid (Figure~\ref{fig:design}).
%

\bpstart{QualCard Structure} We define two states for each QualCard: entry-point and expanded (R2), as shown in Figure~\ref{fig:design}. In its entry-point form, a metric card only shows a main measures view which caters to the entry-point task(s). 
In its expanded form, a metric card adds three customisable sub-views to answer different subsidiary tasks of the types specified in \texttt{ST1}, \texttt{ST2} and \texttt{ST3}. 
Given that each metric's task sequence encompasses multiple tasks of each type, we designed navigation tabs in each subsidiary view to allow users to toggle between different \texttt{categories}, \texttt{quantities} and \texttt{time} granularities  (R3).
Figure~\ref{fig:gold} shows the mapping from subsidiary tasks in a MSS to a QualCard for the ``Gold Standard Drugs'' metric. 
There are two tabs in the categorical sub-view that correspond to the two entries in the \texttt{categories} key on line $23$.
Similarly, there is only one tab on the quantitative sub-view and two tabs in the temporal sub-view, which correspond to elements of the  \texttt{quantities} and \texttt{times} keys, respectively. 
A similar mapping can be found between the mortality MSS in Figure~\ref{fig:mss}$a$ and the sub-views of the mortality QualCard in Figure~\ref{fig:design} (labeled $b$, $c$, and $d$).  

\bpstart{Visual Encoding} We iterated over design alternatives for visual encodings (i.e., mark types) to define preset defaults for each QualCard sub-view.
Alternatives were drawn from  theory~\cite{showme, elshehaly2018taxonomy} and practice~\cite{ft}.
We elicited inputs from front-line analysts to: 
\textit{(a)} set defaults for the visual encodings, and \textit{(b)} design a minimalistic set of GUI elements that support interaction. 
For the main view of a metric card, 
bar charts were strongly advocated by our stakeholders. 
This was clear in the designs collected in our co-design activities (Figure~\ref{fig:activ1}).
Other chart types including funnel plots~\cite{gale2006funnel} and run charts~\cite{perla2011run} are commonly used in practice to report on audit metrics and were also reflected in the co-design sketches. 
Analysts acknowledged that while funnel plots can help visualise institutional variations for a metric, they do not answer questions of what is happening in a specific unit over time. 
Run charts are a common way of process monitoring over time, where the horizontal axis is most often a time scale and the vertical axis represents the quality indicator being studied~\cite{perla2011run}.
We use the same definition for the x- and y-axes of the main view and provide a choice between line and bar mark types. 
Generally speaking, analysts favored bars to line charts.
One PICU consultant stated \textit{``If you give me a bar chart, I don't need to put on my glasses to notice a month with unusually high mortality''}.
This is in line with co-design results reported in~\cite{loorak2015timespan}. 

The choice of mark is specified in the MSS using the \texttt{mark} and \texttt{chart} keys in addition to a logic that is inferred from the \texttt{yaggregates} key. 
To illustrate the latter, the example shown in Figure~\ref{fig:mss}$a$ tells the engine that there are different aggregation rules for the main measures in this metric. 
The first two measures represent counts, while the third is a running average. 
The QualDash engine handles this by generating a dual axis bar chart with differently colored bars representing the two count measures and a line representing the running average. The engine does not allow more than two different rules in \texttt{yaggregates}. 
The \texttt{chart} key tells the QualDash engine which type of bar chart to use for each metric. Supported chart types include stacked (Figure~\ref{fig:gold}) and grouped (Figure~\ref{fig:mss}$b$) bars. Small multiples of simple bar charts were also supported in early iterations of the QualDash prototype, but were deemed less preferable during co-design activities, due to their small size within QualCards. As an alternative, metrics can be rendered in multiple QualCards, effectively using the cards as ``multiples''.  

To set defaults for subsidiary views, we showed different alternatives to analysts and asked them to express their preferences for categorical (ST1), quantitative (ST2), and temporal sub-views (ST3).
To support \texttt{ST1}, a sub-view encodes each of the \texttt{categories} that can be used to break down record groups.  
We use an interactive pie chart for this sub-view (Figure~\ref{fig:design}$b$). 
This design choice is motivated by two factors: 
\textit{(i)} a preference expressed by our analysts to distinguish the categorical domain of these measures from the temporal domain that forms the basis of bar charts in other sub-views (this is in line with best practices for keeping multiple views consistent~\cite{qu2017keeping} ); and
\textit{(ii)} the prevalence of pie chart use in healthcare QI reports which leverages users' familiarity. 

The metric card structure facilitated the discussion around modular interaction design decisions. 
Subsidiary views are linked to the entry-point view, in a manner that is specific to the view's corresponding tasks.
Brushing a pie chart in a \texttt{categories} sub-view highlights the distribution of the corresponding category over time in the main measures view. The converse is also true, in that brushing one or more month(s) in the bar chart causes the pie chart to respond and display the distribution of categories within the brushed cohort. 
Consider, for example, the expanded mortality QualCard shown in Figure~\ref{fig:design} (right).
By highlighting a month with a relatively low number of discharged patients and a low number of deaths (May), one can see that 51\% of patients seen leaving the hospital in that month were recovering from surgery, only 4\% had bronchiolitis, and the remaining patients had other reasons for admission. These percentages are displayed on the pie chart upon mouse hover.
This design bears some similarity to Tableau's part-to-whole analysis using set actions~\cite{setaction}. 

The {\tt quantities} sub-view contributes additional measures that complement the main measures in the entry point view (\texttt{ST2}). 
We use a bar chart for this sub-view and extend the same color palette used for the main view to the measures shown in the {\tt quantities} sub-view. 
To avoid excessive use of color in a single QualCard, we limit the number of measures shown in the entry-point view to a maximum of five measures, and the number of tabs in the {\tt quantities} sub-view to five tabs. 
For metrics that require larger numbers of measures to be displayed, we split them into two or more QualCards and include the appropriate combinations of measures in each.  
Extending the color palette means that the colors used in the entry-point view are not repeated in the {\tt quantities} sub-view.
This design decision was made to avoid any incorrect association of different measures within one QualCard. 
Highlighting a bar in the main view emphasizes the bar corresponding to the same time in this sub-view and adds translucency to other bars to maintain context (Figure~\ref{fig:design}$c$).

The \texttt{times} sub-view adds temporal contexts (\texttt{ST3}). 
In an early design, this view presented a two-level temporal aggregation to facilitate comparison of the same quarter / month / etc. across years. 
This early design used a small multiples view that drew small line charts in each multiple linking the same time unit across years. 
A button enabled users to toggle between this alternative and a traditional multi-series line chart. 
During an evaluation phase (described in Section~\ref{sec:deploy}), participants argued against the small multiples view and preferred to navigate between multi-series line charts of different temporal granularities. 
This co-design outcome is depicted in Figures~\ref{fig:design} and~\ref{fig:gold} which display daily and monthly aggregates, respectively. 
The number of years shown in this view is determined by the \texttt{tspan} key. 
 
 In addition to visual encodings, some textual information is displayed in a QualCard. 
The \texttt{metric} MSS key specifies the title of the QualCard and the \texttt{desc} key specifies  QualCard description that is displayed in a tooltip  upon mouse-hover on the title. 
Figure~\ref{fig:gold}$a$ shows information on the visualisation technique used in the main measures view and indicates the quality of the metric's underlying data by listing the number of missing / invalid values out of the total number of records. 
The area delineated in Figure~\ref{fig:gold}$b$ displays the number of selected records out of the totals in each of the displayed measures. It also includes a ``clear'' button that clears any existing selection. Hovering the mouse on any tab in the sub-views pulls a description of the corresponding data field from the data dictionary and displays it in a tooltip.  

\bpstart{GUI Interactions}
Users can customise the measures shown in each sub-view (R3) via GUI buttons (Figure~\ref{fig:gold}$d$), which allow flexible addition / removal of measures (i.e. tabs) in the categories and quantities sub-views and managing time granularities in the times sub-view. An example  is shown in Figure~\ref{fig:design}$e$ for modifying the categories displayed. 
Additionally, a grey area that outlines a QualCard acts as a handle that can be used to drag and reposition the card on the grid or expand it, when double-clicked. 
A ``control panel'' is included in each QualCard (Figure~\ref{fig:gold}$c$), which contains buttons to download the visualisations shown on the card, and export selected records to a tabular view. 
These exports support the creation of reports and presentations for clinical governance meetings (R6). Finally, a button to expand the QualCard was added to complement the double-clicking mechanism of the QualCard's border. This button was added to address feedback we received in evaluation activities, described in Section~\ref{sec:deploy}. 

\bpstart{Dashboard Layout} We allow users to toggle between \texttt{1x1}, 
\texttt{2x3}, and \texttt{2x2} grid layouts to specify the number of QualCards visible on the dashboard. 
Selecting the \texttt{1x1} layout expands all QualCards and allows users to scroll through them, as one would a PDF slide presentation.
The other two layouts display QualCards in their entry-point form and use screen real estate to render them in either a 2x2 or a 2x3 grid. The engine does not set limits on the number of QualCards to be rendered. 

\section{Evaluation and Deployment}
\label{sec:deploy}

We conducted a series of off-site and on-site evaluation activities to validate the \textbf{coverage}  and \textbf{adaptability} of QualDash.
We define coverage as the ability of the QualCards to be pre-configured (R1) or expanded (R2) to cater to the variety of tasks identified through our task analysis. 
Adaptability is the ability of the QualCards to support different lines of inquiry as they arise (R3).
Additional off-site activities aimed to predict the usability of the generated dashboards; and through on-site observations we sought to establish evidence of the usefulness of QualDash in a real-world healthcare QI setting. 
In this section, we describe the process and outcomes of 62 contact hours of evaluation.
These activities were distributed across our project timeline, with each activity serving a summative evaluation goal, as defined by Khayata et al~\cite{khayat2019validity}.

\subsection{Off-site Evaluation}
We used qualitative methods to establish the coverage and adaptability of the QualCard and the MSS. 
For this, we conducted three focus group sessions with hospital staff. 
We elicited the groups' agreement on the level of coverage that the QualDash prototype offered for the tasks, identified through interviews, and ways in which it could be adapted to new tasks generated through the groups' discussion.

The first focus group session featured an \textit{active intervention with a paper prototype}~\cite{lloyd}. 
The intention was to free participants from learning the software and rather set the focus on the match between the  task sequences, the skeletal structure of the QualCard and the default choice of visualisation techniques in each of its views. 
Appendix C in our supplementary material details the activities of this session and provides a listing of the artefacts used. 
Participants were divided into three groups.
Each group was handed prototype material (artefacts A, B and C in Appendix C) for a set of metric cards. 
For each card, the paper prototype included a task that was believed to be of high relevance,
screen printouts of metric card(s) that address the task, and a list of audit measures that were used to generate the cards. 
A group facilitator led the discussion on each metric in a sequence of entry-to-subsidiary tasks, and encouraged participants to change or add tasks, comment on the data used, and sketch ideas to improve the visualisations. 

After this first session, two sessions featured a \textit{think-aloud protocol}~\cite{rogers2011interaction} with participants in two sites. One of these two groups had engaged with the paper prototype, but both groups were exposed to the dashboard software for the first time during the think-aloud session. 
In each session, we divided participants into groups, where each group consisted of 1-2 participants and a facilitator using one computer to interact with the dashboard. The facilitators gave a short demo of the prototype. 
As an initial training exercise, the facilitator asked participants to reproduce a screen which was printed on paper. 
This exercise was intended to familiarize participants with the dashboard and the think-aloud protocol. 
Following this step, the facilitators presented a sequence of tasks and observed as participants took turns to interact with the dashboard to perform the tasks. 
Each session lasted for 75 minutes and we captured group interactions with video and audio recording. 

Seventeen participants took part in the sessions, including information managers, Pediatric Intensive Care Unit (PICU) consultants, and nurses. 
We analyzed the sketches, observation notes, and recordings from all sessions and divided feedback into five categories: 

\begin{itemize}[noitemsep,nolistsep,leftmargin=*]
    \item A \textit{task-related} category captured the (mis)match between participants' intended task sequences and view compositions supported in the dashboard (R2). 
    \item A \textit{data-related}  category captured comments relating to the data elements used to generate visualisations. This was to assess our findings ($\mathbf{F1 - F3}$) regarding data types included in the structure of the MSS (R1, R2).   
    \item A \textit{visualisation-related} category 
    captured feedback on the choice of visual encoding in each view. 
    \item A \textit{GUI-related} category captured comments made on the usability of the interface.
    \item An \textit{``Other''} category reported any further comments. 
\end{itemize}


The three sessions and follow-up email exchanges with clinicians resulted in a total of 104 feedback comments. 
Task- and data-related feedback constituted 22\% of the comments.  
Data-related feedback at this stage of evaluation focused on issues like data validation and timeliness, rather than on the choice of data elements used to generate the QualCards.
This focus shifted later when we introduced QualDash into the sites, at which point our evaluation participants took interest in accurately specifying the data elements used to generate each card. 
Nonetheless, our off-site activities captured a number of comments regarding aggregation rules. For example, it was noted that Standardised Mortality Ratio (SMR) should be displayed as a cumulative aggregate. 

For task-related comments, we captured feedback in which participants noted a (mis)match between their tasks and the MSS. 
Participants requested  adaptations that were in most cases supported by the MSS. 
One example is for the call-to-balloon metric, where clinicians wanted to 
include a monthly breakdown of patients who did not have a date and time of reperfusion treatment stored in the audit dataset. 
They explained that this would allow an investigation of whether STEMI patients admitted in a time period did not receive the intervention at all; or they did receive it but data was not entered in the audit. 
Accommodating such a request was done by adding a measure labeled as \textit{``No PCI data available''} to the call-to-balloon MSS; for which we selected records having a diagnosis of \texttt{Myocardial infarction (ST elevation)} and a value \texttt{NA} as date of reperfusion treatment. 

Additional tasks were collected throughout the activities. 
Requests were further categorised into customisation issues which could be addressed by simply modifying the corresponding MSS, and design issues, which were addressed in a subsequent design and development iteration. 
An example of the latter is a task that was requested by two PICU clinicians and required adding new MSS functionality. This task enquired about the last recorded adverse event for different metrics.

Visualisation and GUI-related feedback constituted 21\% of the collected comments. 
These comments were largely positive and included a few suggestions to improve readability (e.g., legend position,
size of labels, etc.).
One participant commented: 
\textit{``People approaching the visualisation with different questions can view consistent data subsets and that’s good, because people will try different [sequences] but they’re getting the same answer so this gives us a foolproof mechanism''}.

In addition to this qualitative feedback, we predicted the usability of the dashboards by administering a System Usability Scale (SUS) questionnaire~\cite{brooke1996sus} at the end of each think-aloud session. Participants completed the questionnaire after completing the tasks. Usability scores from the two participant groups were 74 in the first session and 89.5 in the second session, which indicates very good usability.


\bpstart{Enhanced MSS and QualCard}
To provide textual information about adverse events, 
we added a key in the MSS to specify an \texttt{event} which specifies the type of adverse event that clinicians are interested to learn about for the metric.
We further capture the event's \texttt{name}, \texttt{date}, a \texttt{desc} field which describes the event in plain text, and an \texttt{id} field which points to a primary key in the data that identifies the record involving the last reported incident.
This information gets appended as text to the QualCard description tooltip which appears upon mouse hovering the QualCard's title (see Appendix D for more details).  



\subsection{Usefulness of Dashboards in Deployment}

We conducted installation visits at the five hospitals to deploy the QualDash software. 
Prior to each visit, a  hospital IT staff member helped us by setting up a server virtual machine, to which we were granted access via remote desktop. This made it possible for us to access both the client and server on the same physical computer within each site. During each visit, a staff member downloaded raw audit data from the audit supplier's web portal and passed it to the pre-processing R scripts, which in turn fed the  data into QualDash. 
We ran validation tests to ensure that the data were displayed correctly. Through this process, we realised that field headers for the MINAP audit were not unified across sites, due to different versions of the audit being used. 
The QualDash architecture allowed this adaptation by modifying the field names in the config files to match the headers used in each site. 
This adaptation process took approximately 30-60 minutes in each site. 

Following installation visits, we held a series of 10 meetings with clinical staff and data teams in the sites, with the aim of collecting evidence of QualDash's ability to support things like data timeliness (R4), quality (R5) and perceived usefulness of the dashboards' functionality (R1, R2, R3, and R6).
The remainder of this section summarises the evidence we collected along these criteria.

\subsubsection{Support for Data Timeliness}


The client-server architecture of QualDash allows data uploads to take place from any computer in the hospital that has R installed. 
This process was perceived as intuitive by the majority of audit coordinators, who were in charge of data uploads. 
Of the five consultant cardiologists and four PICU consultants that we met, three explained that a monthly data upload was sufficient for their need; while others explained that they prefer uploads to be as frequent as possible. 
One PICU consultant explained that if QualDash was to be updated every week, this would allow them to keep a close monitoring of ``rare events'' such as deaths in unit and accidental extubation.
From the audit coordinators' perspective, monthly uploads were decided to be most feasible to allow time for data validation before feeding it into QualDash. 
One audit coordinator agreed with PICU consultants at her site to perform weekly PICANet uploads. 
In another site, one IT member of staff explained that they would run a scheduled task on the server to support automated monthly MINAP uploads. 
In the general case, however, data upload schedules have been ad hoc and have been driven primarily by need.

All stakeholders appreciated that QualDash allows the right level of flexibility to support each site's timeliness requirements (R4). 
One PICU audit coordinator explained that she was keen on uploading data into QualDash to obtain information which she needed to upload into a database maintained by NHS England to inform service commissioning. This is typically done every three months. 
She explained that the process of extracting data aggregates to upload into this database used to take her up to two hours. However, with the use of QualDash she was able to perform this task in just 10 minutes.

\bstart{Response to COVID-19} One cardiologist (and co-author of this paper) requested to ramp up MINAP data uploads to a daily rate at their site. 
This was in response to early reports of STEMI service delays in parts of the world during the COVID-19 pandemic~\cite{tam2020impact,Garcia27259}. The cardiologist highlighted the need to monitor the call-to-balloon metric card during this time to detect any declines in the number of STEMI admissions (in cases where patients are reluctant to present to the service) and the number of patients meeting the target time. 
This request was forwarded to audit coordinators on the site, who in turn assigned the role of daily validation and upload to the site's data team.  

\subsubsection{Safe Interpretation: A Case Study}
\label{sub:gold}

One of the main motivations behind the client-server architecture and the use of MSSs 
is to ensure that users looking at a specific performance metric share a common understanding of how their site is performing on the metric before their data gets pushed into the public sphere through a national annual report.
Safe interpretation of information in this context relies on capturing patients' care pathways  (R1)  which determine a patient's eligibility for inclusion within that metric. 
The MSS and corresponding metric structure of QualDash allowed for focused discussions with stakeholders at different sites. 
We reflect here on one particular metric called ``Gold Standard Drugs'' (Figure~\ref{fig:gold}), to demonstrate QualDash's support for safe interpretation. 

The Gold Standard Drugs on discharge metric captures the number of patients who are prescribed correct medication upon being discharged from a cardiology unit in a hospital. 
Early co-design activities showed that the main task that users sought to answer for this metric was: 
\textit{What is the percentage of patients discharged on correct medication per month? }
Co-designers indicated that there are five gold standard drugs that define what is meant by ``correct medication''. These include betablocker, ACE inhibitor, Statin, Aspirin and P2Y12 inhibitor. 
Subsidiary tasks for this metric include investigating months with outlier proportions of patients receiving gold standard treatment, for those months, users asked questions such as: Which medication was mostly missing from the prescriptions (\texttt{ST1})? Did the case mix have more patients that were not fit for the intervention (\texttt{ST1})? What was the average weight of patients (\texttt{ST2})? How does the number of prescriptions compare to the same month last year (\texttt{ST3})?

Upon deploying QualDash in one of the sites, two cardiologists pointed out that this metric should not capture the entire patient population but should rather focus on patients eligible for such prescription. 
They explained that eligible patients can be determined from the patients' diagnosis but there was uncertainty about which diagnoses should be included. 
In response to this, we removed the corresponding QualCards from the MINAP dashboards while our team further investigated the patient inclusion criteria for this metric. 
The flexibility of metric card removal was especially beneficial in this case to avoid inaccurate interpretation. We removed the MSS from the configuration files. The remaining parts of the dashboard were not affected. 

Upon further investigating this metric with audit suppliers, we learned that there are only two patient diagnoses that establish eligibility to receive gold standard drugs: ``Myocardial infarction (ST elevation)'', and ``Acute coronary syndrome (troponin positive)/ nSTEMI''. We updated the MSS accordingly and added this QualCard back. 
Here, a known limitation of the MSS design resurfaced when specifying the \texttt{yfilters} for patients who did not receive all five drugs. 
This group presented a composite expression that is not currently supported in QualDash which can be formulated as: 
\texttt{\textcolor{blue}{included\_patient}:= $((betablocker = FALSE)\mid\mid(apsirin = FALSE)\mid\mid(statin = FALSE)\mid\mid(ACEInhibitor = FALSE) \mid \mid (P2Y12Inhibitor = FALSE) )\And(finalDiagnosis \in [$`Myocardial infarction (ST elevation)',`Acute coronary syndrome (troponin positive)/ nSTEMI'$]).$}
To support this composition of $\And$ and $\mid \mid$ operators, we offloaded part of the calculation to a pre-processing step. 
Namely, we added a line in the R script to pre-calculate a field called $missingOneDrug$ which captures the first term of the composition. This simplified the filter to: 
\texttt{\textcolor{blue}{included\_patient}:=} $(missingOneDrug)\And(finalDiagnosis \in [$`Myocardial infarction (ST elevation)',`Acute coronary syndrome (troponin positive)/ nSTEMI'$]). $
The latter fitted nicely in our MSS as shown in Figure~\ref{fig:gold}.
We conclude from the case of drugs on discharge that QualDash's process for MSS configuration enables the management of individual QualCards in different sites. This allows time for the dialogue around the correct data definitions and to verify them from the supplier before pushing the cards back into the sites, to ensure safe interpretation of visualisations.

\subsubsection{Perceived Usefulness: A Case Study}
To investigate the perceived usefulness of QualDash, we present here the case of the mortality metric in the PICUs. 
For this metric, early analysis from our interviews and co-design activities revealed a sequence of tasks that begin with two main questions: 
T1: What is the trend of risk-adjusted standardised mortality rate (SMR) over time? 
T2: What is the raw number of deaths and alive discharges per month? 
From these two entry point tasks, a sequence of subsidiary tasks investigates the case mix (\texttt{ST1}). 
One co-design participant explained the importance of comparing death counts with the same month last year (\texttt{ST3}) as it gives an indication of performance in light of seasonal variations of the case mix.
Additionally, an interviewee explained the importance of considering relevant measures in the same context (\texttt{ST2}) such as the average PIM score (Pediatric Index of Mortality) explaining that \textit{``you say, okay you’ve had 100 admissions through your unit, and based on PIM2 scoring, we should not expect worse than five deaths.''}

To support these tasks, we designed a MSS that captures alive and deceased patients on a monthly basis (Figure~\ref{fig:mss}). 
We added the SMR measure which is aggregated as a running monthly average.
Sub-views include primary diagnosis (for case mix) and monthly averages of PIM score.
Adaptation requests for this card included changes to the x-axis variable, such as deaths to be aggregated in the months in which they occurred or aggregated by date of patient admission. 

When discussing this card with a PICU consultant, she noted \textit{``If we have high SMR, that’s a living nightmare for us, as we would need to investigate every single death. [With QualDash] I can export these deaths and look into the details of these records. That’s very good.
Someone can come in and say your SMR is too high and I can extract all the deaths that contributed to this SMR with 15 seconds effort.''}
As she then looked at the PIM score subview, she noted \textit{``we can also use this to say we need to uplift the number of nurses in [months with high PIM score]. This will be very useful when we do reports and we interface with management for what we need''}.
A clinician in another site observed mortality by ethnic origin (\texttt{ST1}) as he noted \textit{``looking at families of Asian origin, survival rates are better than predicted.''} 
This clinician also noted that the textual display of last recorded event is particularly useful for their team. 
He explained that this information enables him to keep his co-workers motivated for QI by saying something like \textit{``alright, our last extubation was a couple of weeks ago, let us not have one this week.''}

\section{Conclusions \& Future Work }
\label{sec:conc}

We presented a dashboard generation engine that maps users' task sequences in healthcare QI to a unit of view composition, the QualCard. 
Our MSS offers a targeted and more concise specification, compared to expressive grammars like Vega-lite~\cite{satyanarayan2017vega} and Draco~\cite{8440847}, and this is a key reason why QualDash was straightforward to adapt during deployment in the five hospitals.
That made it easy to correct small but important mismatches between clinicians' tasks and our original misunderstanding of them (e.g., the call-to-balloon metric), accommodate new tasks (e.g., for adverse events) and allow site-specific changes. 

The lessons we have learned and factors in QualDash's positive reception lead us to the following recommendations for other design studies: 
\textbf{(a) Trust} in visualised information is enhanced by a level of moderation for dashboard authoring. In healthcare QI, quality metrics have specific patient inclusion criteria that reflect national and international standards. 
MSS configuration files acted as a communication medium between our visualisation team, clinicians and support staff. This allowed for moderated view definitions that ensured safe interpretation of the visualisation. 
\textbf{(b) Modular view composition}, as supported in the QualCards, enables focused communication between dashboard authors, users and system administrators. Comments on QualDash were fed back to us regarding specific QualCards, which enabled refinement and validation iterations to affect localised metric-specific views while leaving the remaining parts of the dashboard intact.
\textbf{(c)  Sequenced rendering} of views, which is materialised by QualCard expansion, provided a metaphor that captured dashboard users' 
task sequences and lines of enquiry pertaining to different metrics. 
Further evidence is required to establish the usability of the MSS as an authoring tool, and for that we have commenced a field evaluation of QualDash in the five hospitals.
The results will be reported in a future paper.

We explored the generalisability of the Qualcard through discussions and demonstrations with clinicians and Critical Care research experts who are outside of the QualDash stakeholder community. 
The idea of custom-tailored visualisation cards that capture tasks sequences was very well received. 
This has led to budding collaborations as demand for this type of adaptable dashboard generation is gaining momentum with the diversity of tasks surrounding COVID-19 data analysis. 
We have received questions about whether we can generate QualCards to support decision makers' understanding of risk factors and vulnerable communities. 
We have also received some questions about the possibility of adding more visualisation techniques, like maps for geospatial data.
In the current version of QualDash, we only support time and population referrer types~\cite{andrienko2006exploratory}. However, based on these discussions, we foresee great opportunities to further develop our engine and support more referrers like geographic location.

Finally, we plan to identify new design requirements for possible transitions across QualCards. 
We expect that the modular nature of the self-contained QualCard will help focus these design decisions to localised areas of the dashboard screen and to specific task sequences.
Such transitions were not found necessary for healthcare QI dashboards in our experience, but may be deemed necessary in other applications. 

\acknowledgments{
This research is funded by the
National Institute for Health Research (NIHR) Health Services and Delivery Research (HS\&DR) Programme (project number 16/04/06). 
The views and opinions expressed are those of the authors and do not necessarily reflect those of the 
HS\&DR Programme, NIHR, NHS or the Department of Health.
}

\bibliographystyle{abbrv-doi}

\bibliography{vis19}
\end{document}